\documentclass{nature}

\usepackage{color} 
\usepackage{footnote}
\makesavenoteenv{tabular}
\makesavenoteenv{table}
\usepackage{amsfonts}
\usepackage{pdfpages}
\usepackage[]{algorithm2e}
\usepackage{amsmath}
\usepackage[T1]{fontenc}
\usepackage{lmodern}
\usepackage[normalem]{ulem}
\usepackage{color,soul}
\usepackage{graphics}
\usepackage{textcomp}
\usepackage{gensymb}
\usepackage{amsmath}
\usepackage{amssymb}
\usepackage[switch, modulo]{lineno}

\usepackage{graphicx}
\makeatletter
\let\saved@includegraphics\includegraphics
\AtBeginDocument{\let\includegraphics\saved@includegraphics}
\renewenvironment*{figure}{\@float{figure}}{\end@float}
\makeatother

\title{Estimation of rheological parameters for unstained living cells}

\author{Kirill Lonhus, Renata Rycht\'{a}rikov\'{a}, Ali Ghaznavi \& Dalibor \v{S}tys}

\begin{document}
\maketitle

\begin{affiliations}
 \item University of South Bohemia in \v{C}esk\'{e} Bud\v{e}jovice, Faculty of Fisheries and Protection of Waters, South Bohemian Research Center of Aquaculture and Biodiversity of Hydrocenoses, Kompetenzzentrum MechanoBiologie in Regenerativer Medizin, Institute of Complex Systems, Z\'{a}mek 136, 373 33 Nov\'{e} Hrady, Czech Republic
\end{affiliations}

\begin{abstract}
In video-records, objects moving in intracellular regions are often hardly detectable and identifiable. In order to squeeze the information on the intracellular flows, we propose an automatic method of reconstruction of intracellular flow velocity fields based only on a recorded video of an unstained cell. The basis of the method is detection of speeded-up robust features (SURF) and assembling them into trajectories. Two components of motion -- direct and Brownian -- are separated by an original method based on minimum covariance estimation. The Brownian component gives a spatially resolved diffusion coefficient. The directed component yields a velocity field, and, after fitting the vorticity equation, estimation of the spatially distributed effective viscosity. The method was applied to videos of a human osteoblast and a hepatocyte. The obtained parameters are in agreement with literature data.
\end{abstract}

\section{Introduction}

\label{intro}
A typical bright-field microscopy experiment is time-lapse recording of a sequence of images. In case of living unstained samples, it is little known about structure of the observed objects. It is usually possible to discriminate a cell from its background, find its nucleus, but not more.\cite{Buggenthin2013} However, the microscopy image is much more complicated and one can see motion of some intracellular structures and movement of small 'particles' inside the cell. These objects are extremely diverse in texture and shape, frequently do not have sharp boundaries, and are mostly too small for identification. 

In this article, we aim to investigate cell rheological and microfluidic properties without any \textit{a priori} information about cell structure or composition. There are approaches aimed specifically at investigation cell flows, e.g.,~\cite{BoquetPujadas2017}, but they require fluorescent labelling and a mathematical model of the studied cell. There are model-free approaches as well. These are based on correlation computations, e.g.,~\cite{Crocker2007}, have a solid mathematical background and, at good conditions and for well-behaved objects, can deliver good results. But these correlation methods suffer from the fact that they cannot distinguish the points and rely on proximity based assignment. As a result, these methods inevitably suffer from error propagation during tracking. Another way is to segment some sufficiently large objects and then track them until they are overlapping, e.g.,~\cite{Rychtarikova2017}. These methods do not suffer from the error propagation so much, but require segmentable entities in the cell image. Even then, the count of followed objects can be too small for flow reconstruction. Moreover, all methods described above do not address the fact that small particles can be susceptible to the Brownian motion. All the methods also often assume that the random component of motion can be safely neglected. 

The main idea of the method proposed here is tracking of identifiable spots inside a cell followed by reconstruction of local properties of media and fields of velocities. This approach is similar to two well-known model-free approaches to the velocity reconstruction such as the Particle Image Velocimetry (PIV)\cite{Melling1997} and the Particle Tracking Velocimetry (PVT)\cite{LUeTHI2005}. After that, the non-linear optimization of minimum covariance, alternating likelihood fitting, enables us to separate the observed motion to components of the Brownian and direct flow, respectively, yielding both rectified flows and local media properties.

\section{Materials and methods} \label{methods}
In order to show capacity of the method, we applied it to microscopic image data from time-lapse experiments on live human cells of lines MG63 and HepG2.

\subsection{Cell sample preparation}
A MG63 (human osteosarcoma, Sigma-Aldrich, cat. No. 86051601) and a HepG2 (human hepatocellular carcinoma, Sigma-Aldrich, cat. No. 85011430) cell lines were grown at low optical density overnight at 37$^\circ$C, 5\% $\mbox{CO}_2$, and 90\% RH. The nutrient solution consisted of DMEM (87.7\%) with high glucose ($>$1 g L$^{-1}$), fetal bovine serum (10\%),  antibiotics and antimycotics (1\%), L-glutamine (1\%), and gentamicin (0.3\%; all purchased from Biowest, Nuaill\'{e}, France). 

During the microscopy experiments, the MG63 cells were maintained in a Petri dish with a cover glass bottom and lid at room temperature of 37$^\circ$C. The HepG2 cells were cultivated in a Bioptechs FCS2 Closed Chamber System at 37$^\circ$C (Tab. 1).


\subsection{Bright-field wide-field videoenhanced microscopy}


The living cells were captured using a custom-made inverted high-resolved bright-field wide-field light microscopes enabling observation of sub-microscopic objects (ICS FFPW, Nov\'{e} Hrady, Czech Republic): The HepG2 line was captured by an older type of microscope (so-called nanoscope, built 2011), whereas the MG63 cell line was scanned using a newer type of microscope (so-called superscope, built 2020). 

The optical path of the both microscopes is very simple and starts by a light emitting diode(s) which illuminate(s) the sample by series of light flashes (synchronized with a microscope digital camera exposure and image saving speed) in a gentle mode and enable the videoenhancement.\cite{Rychtarikova2017} In the case maybe, a light filter is applied to protect the sample from undesirable intensities. After passing through a sample, light reaches a Nikon objective. In the nanoscope, a Mitutoyo tube lens magnifies and projects the image on a high-resolved rgb digital camera. At this total magnification, the size of the object projected on the camera pixel is under the Abbe diffraction limit, i.e. 32 and 23 nm, respectively. The process of capturing the primary signal was controlled by a custom-made control software. In both cases, we performed a time-lapse experiment from a compromise focal plane of the cell. The microscope set-ups differ as written in Table~\ref{Tab1}.

\begin{table}
\footnotesize
\caption{Bright-field wide-field microscopy constructions and set ups.}
\label{Tab1}
\begin{center}
\begin{tabular}{l l l p{3cm}}
\hline
\bfseries Microscope (cell)    & \bfseries nanoscope (HepG2) & \bfseries superscope (MG63)\\
    \hline
\bfseries LEDs &2$\times$ Luminus CSM-360, & 1$\times$ Luminus CFT-90-W,\\

\bfseries &4500 mA (59.625 W) &40\% of max. intensity \\

\bfseries{Light pattern} &light
226.1 ms–dark 96.9 ms & light 0.2 ms-dark 199.8 ms\\

\bfseries Light filters &Edmund Optics, & no\\

 & i.r. 775 nm short-pass, & \\

\bfseries & u.v. 450 nm
long-pass &\\

\bfseries Objective & Nikon LWD 40$\times$, Ph1 ADL, & Nikon CFI Achromat 60$\times$, \\

\bfseries & 1/1.2, N.A. 0.55, W.D. 2.1 mm & N.A. 0.80, W.D. 0.30 mm\\

\bfseries Tube lens &Mitutoyo, 4$\times$ &no \\

\bfseries Camera &JAI, rgb Kodak KAI-16000 chip, & Ximea MX500-CG-CM-X4\\

\bfseries & 4872$\times$3248 px & G2-FL rgb, 7920$\times$6004 px\\

\bfseries Camera Bayer mask & GBRG & BGGR\\ 

\bfseries Camera exposure &293.6 ms (gain 0, offset 300) &0.2 ms\\

\bfseries Pixel size &32 nm &23 nm\\

\bfseries Scanning frequency &0.2 fps & 5 fps\\

\bfseries Experiment length  &2446.869 s &83.2 s \\

\bfseries Cell cultivation &Bioptechs FCS2 Closed & Ibidi µ-dish 35 mm, high\\

\bfseries &Chamber System & glass bottom, DIC lid\\

\bfseries No. of px per cell &(2.137 $\pm$ 0.048) $\times$ 10$^6$ & (5.623 $\pm$ 0.084) $\times$ 10$^5$\\

\bfseries No. of images &473 &416 \\
\hline
\end{tabular}
\end{center}
\end{table}

\subsection{Image Preprocessing}
In order to suppress the image distortions, the microscope optical path and camera chip was calibrated and the obtained time-lapse micrographs were corrected by a radiometric approach described in detail in~\cite{Lonhus}.

The raw images were recorded in the color preserving RGB mode when three intensity values (in the red, green, and blue image channel) are assigned to each image point (pixel). In this colour preserving image representation, four camera pixels are always merged in a way that the resulting number of the RGB image pixels is a quarter (see~\cite{LIL} for details). In other words, the resulting pixel size is doubled, i.e., 64 nm and 46 nm, respectively (cf. Tab.~\ref{Tab1}).  Since all examined feature detectors work on single-channel images, the RGB images were converted to grayscale in the standard way~(0.2989$\cdot R$ + 0.5870$\cdot G$ + 0.1140$\cdot B$, where $R$, $G$, and $B$ are intensities of pixels in the red, green, and blue raw image channel, respectively).\cite{rgb} To eliminate subtle changes in illumination, the images were robustly rescaled to $[0..1]$, after saturating $1\%$ of both the darkest and the brightest pixels simultaneously.

\textit{Prior to} any tracking, the objects of interest (live cells) have to be robustly detected and segmented from image background.
Therefore, we annotated a few (usually 1\%) images from the sequence visually to interpolate contours of the observed cell in the unannotated images. For interpolation of the contours, we used a weighted mean of strings.\cite{Jiang} After contours were interpolated, we applied a non-parametric image deformation registration.\cite{Thirion1998} The obtained displacement field was employed to compensate position shift between the images.

\section{Estimation of intracellular flows}

\begin{figure}
\label{Fig1}
\resizebox{1.0\columnwidth}{!}{\includegraphics{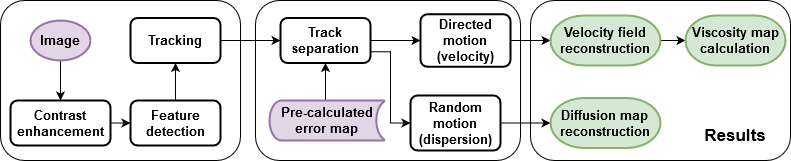}}
\caption{Algorithm of the method for calculation of the viscosity map and diffusion map of the intracellular environment.}
\end{figure}

The algorithm for the estimation of the flows and rheological parameters in the intracellular environment of the unstained cells is showed in Figure~\ref{Fig1} and described in detail in the following subsections. The Matlab codes and the input and output data are available at the Dryad data depository.\cite{dryad} 

\subsection{Feature extraction and tracking}
\label{feat}
There are numerous methods, e.g.,~\cite{Li2008,Latif2019}, for tracking local image features, i.e., feature vectors describing special, well-distinguishable image points. These methods are usually designed to match the same object from different views. Our problem is opposite -- to match different (but similar) objects from the same view. We tested BRISK\cite{Leutenegger2011}, ORB\cite{Rublee2011}, MSERF\cite{Mikolajczyk2005}, KAZE\cite{Alcantarilla2012}, MinEig\cite{Shi1994}, and SURF\cite{Bay2008} image features to estimate their efficacy (Fig.~\ref{Fig2}b; see Sect.~\ref{decomp} for determination of the error in separation of the direct motion from the random walk). The SURF performs the best, followed by the MinEig. The further analysis showed that the SURF output is much more robust to small changes in the image. The SURF method is based on calculation of the Hessian matrix for each pixel of the smoothed (via approximated Gaussian smoothing; a box filter with kernel 9$\times$9 px and $\sigma=1.2$) image separately. The pixels whose matrix determinants were maximal were treated as the 'points'. An image pyramid with 3 scales was further used. The descriptors themselves were oriented Haar wavelets.\cite{Bay2008} 

The next step was to track a point through consecutive frames. To avoid a computationally intensive $O(n^2)$ point match (where $n$ is a number of points in an image), we used a heuristic approach --- the same points in consecutive frames should be nearby. A small, random, subset of ($\sim10$) pairs of consecutive images was used to estimate the maximal point displacement in two images: For each pair of the consecutive frames, we found a median of the minimal distances between each two points. Then, the resulted effective displacement $ED$ was calculated as a mean from all medians of the minimal distances. Finally, we assume that the match between the points is possible if the distance is smaller than $3\cdot ED$. In this way, each point obtained typically 10--15 possible candidates for tracking in the following image and, thus, we effectively reduced feature matching complexity to $O(n)$ and eliminated the long-range matching error.

The tracking process itself is iterative. At each step we classified all detections into two sets: assigned and unassigned. To be assigned, a detection in any track had to fulfil two criteria -- to be spatially close (closer than 3 average offsets) and feature-wise close (the Euclidean distance between the last and the current vector of the track has to be smaller than 1). The unassigned detection created new tracks. The tracks which were not assigned for a longer period than $K$ frames were removed. Since the influence of $K$ on quality of the final result has not been investigated, we used the safest choice of $K=1$. 

\subsection{Decomposition to direct and Brownian motion}
\label{decomp}
The segmented trajectories are sets of points in $\mathbb{R}^2$, usually 10--300 points. We assume that the trajectories exhibit two simultaneous types of motion -- Brownian and direct. As widely accepted (the Einstein model), the Brownian motion of small particles can be described as a Gaussian process with zero mean. To separate the components of motion, we used the minimization of a maximum differential entropy, which for a multivariate normal distribution follows $h(x)\le \frac{1}{2}\log\det \mbox{cov}(\vec{X})$. In this way we proposed a formulation of the separation problem as
\begin{equation}
\label{eq:brown}
\vec{V}_{d} = \underset{\vec{V}\in \mathbb{R}^2}{\mbox{min}} \log |\mbox{cov}(\vec{P}_n - n\vec{V})|, 
\end{equation}
where $\vec{P}_n$ is a position of the tracked point in time step $n$ and $\vec{V}_d$ is the searched velocity. Equation 1 can be also viewed as direct usage of the minimum covariance approach. 

This optimization also gives a corrected (with a compensated drift) set of points from which 'normal' covariance and mean value can be estimated. We chose a nonlinear optimization -- sequential-quadratic programming\cite{Burke1989} -- which, in the vicinity of a current point, iteratively approximates a nonlinear problem by a quadratic one and solves this simpler problem by a QR decomposition. This method is not global and relies on the initial guess. We used the safest guess -- the zero velocity -- which coincides with the null hypothesis. 

In order to verify this approach, we performed the following numerical experiment (simulation): The most straightforward way how to mimic the Brownian motion is the random walk, where the steps are drawn from the Gaussian distribution. The simulation itself has two main parameters: a number of points $N$ in a track and fuzziness $\frac{\sigma}{|\vec{V}|}$, where $\sigma$ is a standard deviation of the Gaussian process $\mathcal{N}$ and $\vec{V}$ is a drift velocity vector. Then the position of the tracked point in time step $(n+1)$ is
\begin{equation}
\label{eq:rwalk}
\vec{P}_{n+1} = \vec{P}_{n} + \vec{V} + \mathcal{N}(0, \sigma).
\end{equation}
After that, for any random walk with drift, it is possible to apply the resulted components of the method of separation of the direct motion from a random walk and evaluate the error $Err=\frac{|\vec{R}-\vec{V}|}{|\vec{V}|}$, where $\vec{R}$ and $\vec{V}$ is the reconstructed and real velocity, respectively.

\begin{figure}
\resizebox{1.0\columnwidth}{!}{ \includegraphics{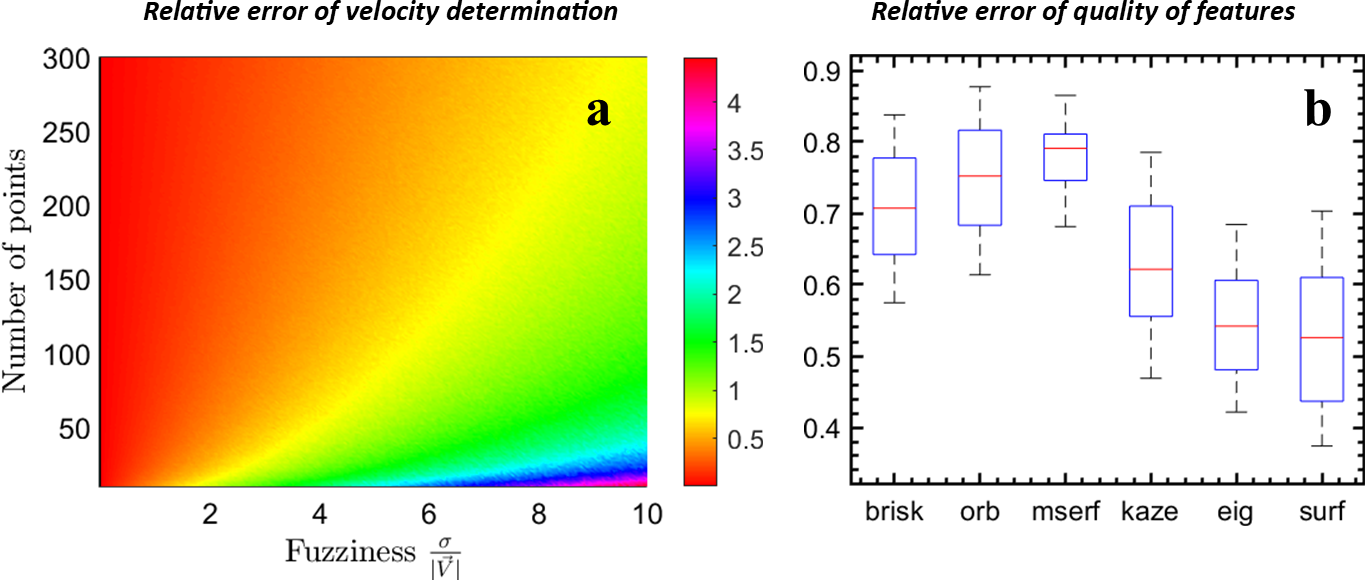} }
\caption{(a) Relative error of velocity determination as a function of number of points in trajectory and ratio between standard deviation $\sigma$ and norm of the velocity $V$. (b) Relative error of quality of features for feature extraction methods.}
\label{Fig2}  
\end{figure}

Using equation~\ref{eq:rwalk}, we simulated numerous tracks varying in the number of time steps (from 8 to 300) and in the fuzziness (from 0.01 to 10 discretized into 500 steps). The data along all 500 trials were averaged and saved as a table (Fig.~\ref{Fig2}a). By a 2D bilinear interpolation, it was allowed to calculate the error of velocity extraction $Err$ from a non-synthetic data. It requires that the velocity is both spatially and temporarily constant (along the given track) and the observed random motion obeys the Gaussian distribution.

If the data variation is not too high ($\sigma/|\vec{V}|<0.1$), we can carry out a reliable (relative error $Err < 0.01$) extraction of the drift velocity from sets of down to 10 points. For a higher number of points, the drift velocity extraction gives a quite reliable estimation even if the standard deviation is much greater than the norm of the drift velocity vector.

Due to absence of the ground truth, there is no way how to evaluate quality of the reconstructed flows. But quality of the tracks can be evaluated as the mean separation error of the tracks. In this way, we compared the different feature detectors, defining that a lower reconstruction error means a better detector (Fig.~\ref{Fig2}b, more above in Sect.~\ref{feat}).

\subsection{Reconstruction and analysis of intracellular flows}
\label{flows}
The velocities were defined for the most of the tracks. Some of the tracks were excluded from the future analysis due to a high separation error (the threshold value was chosen 1). There was no way how to attribute the given velocity to the specific position, because we estimated the drift for the whole trajectory. We assumed that the drift is constant along the observed positions in the trajectory. All tracks' velocities were imprinted in a single global image of the cell.

The particles passing through the same point (in 2D projection) at the same time can exhibit completely different velocities. These velocities have to be separated. Since we calculate velocities along the time window, for each pixel we obtain as many estimations of velocities as length of the time window. From these different estimations of velocities, we can calculate the error of velocity separation $Err$ (see Sect.~\ref{decomp}). In following statistical analysis, we will assign weights to the velocities estimated in this time window. Each of this weight is complementary to the error of separation, i.e., $weight=1-Err$. 

The resulted vector field is sparse. In order to reconstruct it, we used robust splines\cite{Garcia2010a} which minimize the Generalized Cross-Validation (GCV) score. This method was designed to handle the PIV-type data specifically.\cite{Garcia2010}

Eventually, this part of the algorithm produces a global velocity field through the whole image series. In view of the fact that it is not possible to do any real time series analysis, we carried out a quasi-stationary window analysis. The reconstruction was performed on subsets of frames defined by the time window of the size $wsize$ sliding along the whole image sequence. The time window is usually too short to give a reliable reconstruction and, thus, the global flows are used as a guess (with dampened weights) proportional to the ratio between the window size and the total number of images in the series. The resulted velocity field (as a function of the sliding window size) is the closest form how we can approximate the real time dependence of the velocity field.

\begin{figure}
\resizebox{1.0\columnwidth}{!}{ \includegraphics{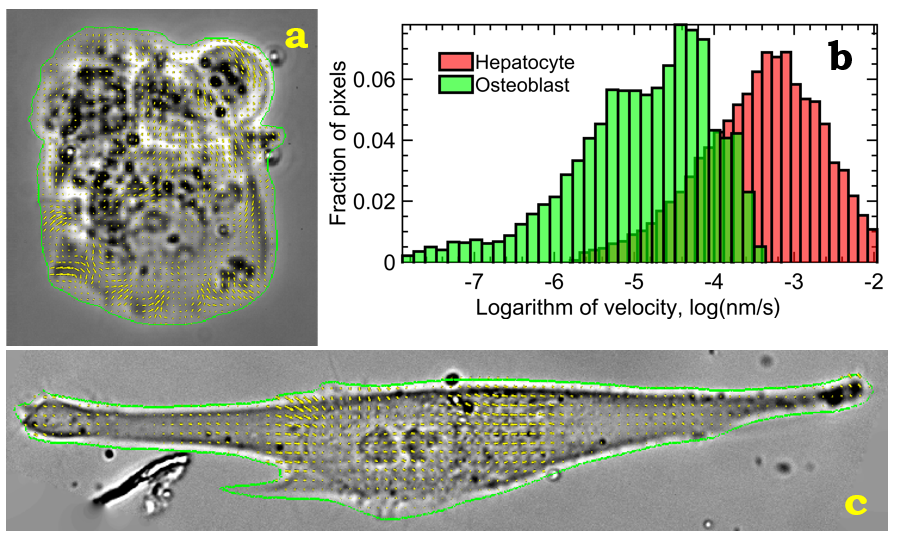} }
\caption{The reconstructed global velocity field for a hepatocyte (a) and osteoblast (c). The corresponding velocity frequency histograms are shown in panel (b).}
\label{Fig3}  
\end{figure}

We applied the method to two types of objects -- a human osteoblast and human hepatocyte observed with bright-field microscopy (see Sect.~\ref{methods}). The main output of the method is a velocity field and distribution  of flow speeds (Fig.~\ref{Fig3}). It is predictable that the intracellular flows in the hepatocyte (a cell with high metabolic activity) are much more intense than in the osteoblast.

\subsection{Diffusion and viscosity estimation}
The velocity is informative enough, but it does not characterize the intracellular medium itself. In order to characterize the structure and composition of the medium, some hydromechanical constants, namely space-resolved diffusion coefficient and viscosity, must be extracted.

\begin{figure}
\resizebox{1.0\columnwidth}{!}{ \includegraphics{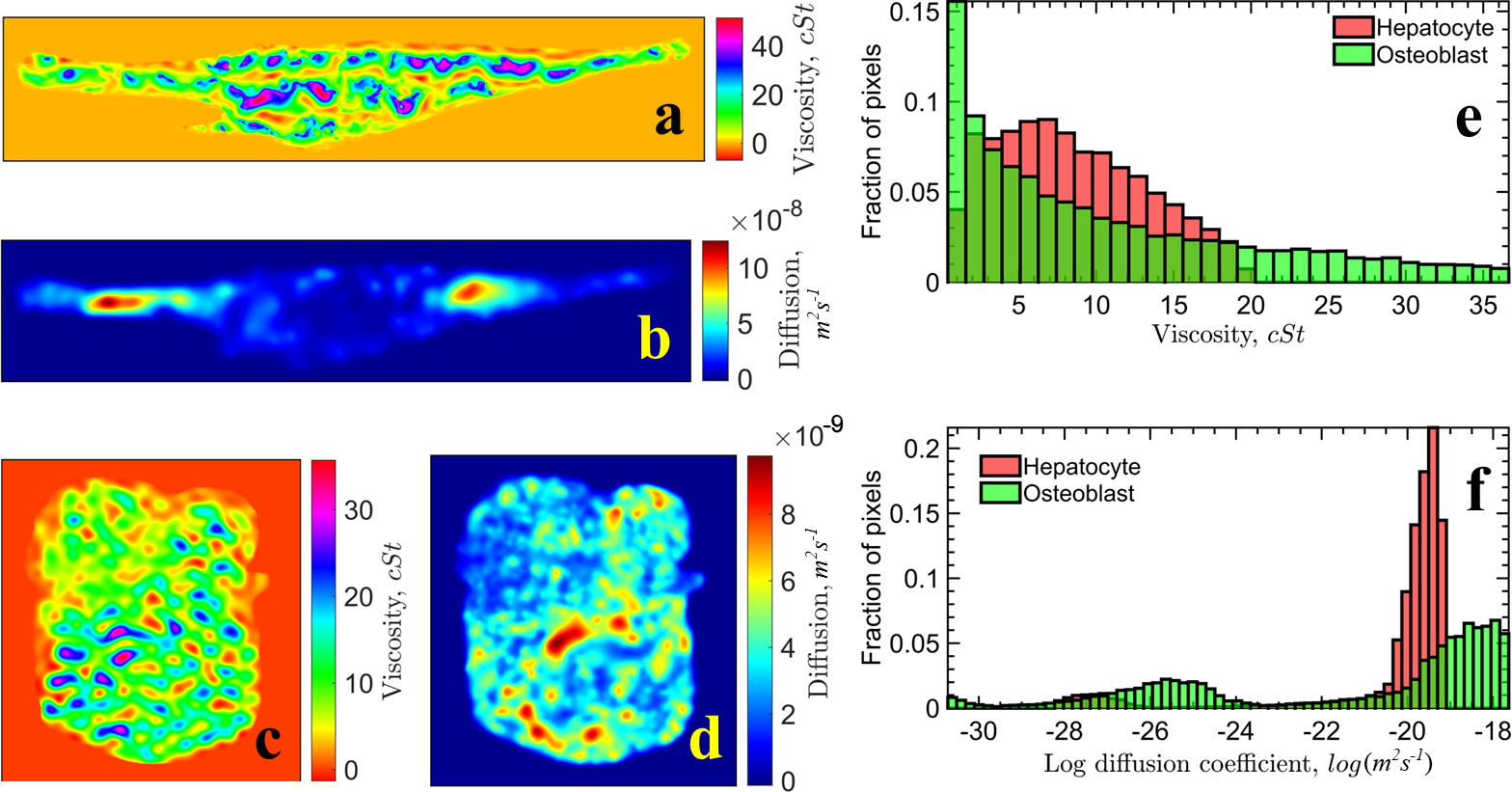} }
\caption{The maps of intracellular effective diffusion and viscosity coefficients for a hepatocyte (c, d) and osteoblast (a, b). The relevant frequency histograms of the viscosity and diffusion coefficients are in panels (e, f).}
\label{Fig4}  
\end{figure}

The separation procedure resulted in the drift-compensated trajectory (see Sect.~\ref{decomp}). The most straightforward way how to estimate the diffusion coefficient is to use the covariance of derivatives in the random walk:
\begin{equation}
\label{eq:diffusion}
D=\frac{1}{4T}<\mbox{diag cov} \frac{d\vec{P}_{n}}{dn}>, 
\end{equation}
where $T$ is the time interval between consecutive images. Due to presence of derivative in equation~\ref{eq:diffusion}, the diffusion coefficient is invariant to the drift velocity as it was supposed to. These diffusion coefficients were computed for all eligible ($Err < 1$) tracks. The field of diffusion coefficients was reconstructed in the same way as the velocity field, i. e., by a spline minimizing the GCV score. The reconstructed diffusion fields and distributions can be seen in Figure~\ref{Fig4}b, c, f. The values of diffusion coefficients are relatively high, presumably because both the active and passive diffusion happen in the same time and are mutually indistinguishable. Essentially, we deal with effective diffusion and, thus, the comparison with classical molecular diffusion coefficients should be done with caution. Since we work with a 2D slice of a 3D volume, the value of the derived diffusion coefficient should be accurate, assuming its isotropy. No additional smoothing of the final data was used, except removing 5\% of points with the least and most intensities, respectively, before reconstruction (to eliminate possible influential errors). 

Estimation of the viscosity coefficient is less model-free and based solely on the quasi-stationary velocity field. The kinematic viscosity\cite{Rossi2015} can be found from the vorticity equation for an incompressible, isotropic, Stokesian fluid in 2D as
\begin{equation}
\label{eq:viscosity}
\nu = \frac{d\omega}{dt}\cdot \frac{1}{{\nabla}^2 \omega},
\end{equation}
where $\omega=\nabla\times\vec{V}$ is the vorticity of the velocity field. One issue of this approach is a high, namely the 3rd, order of derivatives in the spatial domain. This leads to the fact that the calculations will be thus over-susceptible to small errors. The second issue is presence of the time derivative that is absent in the results because the analysis is quasi-stationary and the intracellular flows thus depend on the time window. The window, which we used in the analysis and was the closest to zero, was $7$. With decreasing size of the time window, the absolute error is increasing due to less rich statistics. For all windows from 7 to 71 images (only odd numbers are valid as the window size), we calculated the mean velocity field and mean time derivative. The distances between windows $[w, w+wsize]$ and $[w+1, w+wsize+1]$ were assumed 1 frame. But this is strictly true only for $wsize=0$ and diverges with increasing size of $wsize$. Thus equation~\ref{eq:viscosity} was applied to each window and then extrapolated to $wsize=0$. Due to the higher-derivative noise, the ordinary linear fitting was not sufficient for the extrapolation. Therefore, we had to apply a robust linear fitting\cite{Holland1977} with bi-square weights, which gave stable results without necessity of any additional data smoothing (Fig.~\ref{Fig4}a, d, e). 

The obtained values of viscosity are in agreement with some literature data.\cite{Kuimova2009} Nevertheless, some literature sources report much lower viscosities.\cite{Parker2010} It can be explained by the fact that the definitions of viscosity at the microlevel are very vague, the relevant values of viscosity then depend frequently on the method of their acquisition, and, thus, the real values of viscosity can vary. Again, we work with a single plane of a 3D object and, thus, diffusion and convection along the z axis is neglected. Therefore, it is more correct to call the variable derived here as effective viscosity.

\section{Discussion}
\label{disc}

In this paper, we deal with the total, complex, evaluation of the intracellular flows but the origin of the intracellular flows remains an open question. We can observe visually that these flows do not coincide with specific object motions. In most cases, it is nearly shapeless disturbance in the intracellular medium which is moving, sometimes we deal with small particles or vesicles. We do not speculate nature of these objects or nature of their motion and rather try to analyse it.

The main assumption for the flow analysis is that the tracked entities are driven by two forces -- the Brownian and direct motion -- which are related to both some global intracellular flow (if exists) and a specific locomotion. The reconstructed flows seem not to be any consequence of the changes in the cell borders but rather some intrinsic phenomena. In an effort to interpret the results from the biological point of view, we chose two very mutually different kinds of cells -- osteoblast (bone cell, low mobility, low metabolism) and hepatocyte (liver cell, medium mobility, intense metabolism). 

There is no literature data about such intracellular velocities but, at least, their distributions follow a general meaning of cell physiology -- more intense metabolism coincides with a higher mean and median of the velocity (Fig.~\ref{Fig3}). To compare the results of the described method with other methods, we estimated the hydromechanical parameters of the intracellular medium. The proposed separation procedure yields a local standard deviation of the random walk-like process, which can be naturally converted to a effective diffusion coefficient (Fig.~\ref{Fig4}b--c). But any comparison with other results is complicated, because most of the diffusion coefficients are determined for molecules but we presumably observe motion of larger intracellular structures. 

The obtained effective diffusion coefficients are in the range 10$^{-10}$--10$^{-8}$ m$^2$ s$^{-1}$ and correspond to values for particles in liquids.\cite{He2015} The resulted coefficients may be related to both active and passive diffusion. Namely, the diffusion map of the osteoblast is very inhomogeneous but this has no relation to the velocity distribution (cf. hepatocyte in Figures~\ref{Fig3}b and~\ref{Fig4}f). In the osteoblast's interior, there are two sites with very high diffusion coefficients (likely active diffusion) and the central region of low diffusion. This central region roughly corresponds to the position of nucleus (as guessed from the typical structure of osteoblasts; in the raw images, nucleus is not observed at all, because the microscope was focused on the cell surface). 

The kinematic viscosities for both cells are in the range 5--50 cSt, which is comparable with palm oil and other viscous substances. The dispersion of viscosity for the osteoblast is much higher, but there is no much explanation for this. The resulted viscosity fields are quite noisy, since the numerical estimation of the 3rd derivative is a quite sensitive process. Surprisingly, the values are meaningful even without advanced smoothing. However, for in-depth analysis of the maps, we definitely need a more sophisticated processing. However, we observe only a planar slice of a 3D system and the equations here were derived for 2D. Thus, the obtained viscosity is rather effective than true, physical. Nevertheless, it is possible to compare the values of this quasi-viscosity between similar experiments; or do extensive validation and find a correction factor to obtain real kinematic viscosity and conditions, where such a explicit continuous mapping exists. Despite all the facts, a single plane derived viscosity has a reasonable scaling and, thus, may be compared with other viscosities, but with caution. 

The main advantage of the intracellular rheology estimation method described in this paper is its simplicity. As seen in this paper, the algorithm works with time-lapse image series of unstained living cells in any bright-field microscope (we show independent results for time-lapse series from two different bright-field microscopes, see Section~\ref{methods}). Nevertheless, let us note that this method can be applied in analysis of fluorescent image data. If applied, the complete analysis of flows in the stained living cells would be simplified compared to the bright-field data (due to a lower number of the possibly detected and tracked points and their identification). However, the biological relevance of such results is debatable, since the fluorophores can be cytotoxic and can completely change cell metabolismus and dynamics. Thus, only autofluorescence plays an important and obvious role in interpretation of the intracellular dynamics.

Also, the algorithm described here does not require any \textit{a priori} given constant or assumptions about processes in the sample. Moreover, we have studied only one semi-tomographic slice of an active, unstained, 3D object, which can make the biologically relevant interpretation even more tricky. At least we know that the described values are sufficiently stable and, therefore, can be used for cell characterization. The conducted experiments are rather illustrative than explorative. We have not so far dealt with linking the results to biology but, compared with the literature, e.g.~\cite{Parker2010,Puchkov2013,Dench2016}, they seem to be promising.




\section{Conclusions} \label{conc}
Better understanding of a cell behaviour is one of the major task of modern biology and key to very important technologies such as growing artificial tissues and organs, or fighting against cancer. In such challenging tasks, biologists will need as many reinforcements as possible. And this method, among others, is aimed to bring physicists, data scientists, and mathematicians to life sciences; and make a shortcut between classical, wet, biology and formidable machinery of modern data explanatory analysis and machine learning. Therefore, the approach is quite minimalistic. For application, one needs only a video with living cells and knowledge of a camera sensor geometrical size. The outputs of the method are physically understandable and interpretable parameters. But the origin of such flows and the overall cell fluid dynamics is a different story, and, hopefully, will be solved in the meantime.

\vspace{5mm}

\small{This work was supported by the Ministry of Education, Youth and Sports of the Czech Republic project CENAKVA (LM2018099), GAJU project 013/2019/Z, and from the European Regional Development Fund in frame of the projects Kompetenzzentrum MechanoBiologie (ATCZ133) and ImageHeadstart (ATCZ215) in the Interreg V-A Austria-Czech Republic programme. The authors would like to thank Petr Mach\'{a}\v{c}ek (Image Code company, Brloh, Czech Republic) for software development, Petr Tax (Optax company, Prague, Czech Republic) for custom microscope development, and Miroslav Slivon\v{e} (a USB student) for the HepG2 microscopy data acquisition.}


\bibliographystyle{naturemag}
\bibliography{Lonhusetal}

\end{document}